\documentclass[aps,prl,floatfix,reprint,twocolumn,superscriptaddress]{revtex4-1}
\usepackage{amsmath}
\usepackage{graphicx} 
\usepackage{xcolor,colortbl} 
\usepackage{xurl}
\usepackage{hyperref}

\newcommand{\E}{\mathcal{E}}

\newcommand{\ket}[1]{| \mathrm{#1} \rangle}
\newcommand{\bra}[1]{\langle \mathrm{#1} |}

\def\EX{E_\mathrm{X}}
\def\Edark{E_\mathrm{D}}
\def\Ephoton{E_\mathrm{C}}
\def\Obiexc{\Omega_{\mathrm{XD}}}
\def\ORabi{\Omega_{\mathrm{Rabi}}}

\begin{document}
\title{Biexciton-polariton coupling mediated by dark states}
\author{G.~Fumero}
\email{giuseppe.fumero@gmail.com}
\affiliation{Associate, Nanoscale Device Characterization Division, National Institute of Standards and Technology, Gaithersburg, MD 20899, USA}
\affiliation{Department of Physics and Astronomy, West Virginia University, Morgantown, WV 26506-6315, USA}
\author{J.~Paul}
\affiliation{Associate, Nanoscale Device Characterization Division, National Institute of Standards and Technology, Gaithersburg, MD 20899, USA}
\affiliation{Department of Physics and Astronomy, West Virginia University, Morgantown, WV 26506-6315, USA}
\author{J.~K.~Wahlstrand}
\affiliation{Nanoscale Device Characterization Division, National Institute of Standards and Technology, Gaithersburg, MD 20899, USA}
\author{A.~D.~Bristow}
\email{alan.bristow@mail.wvu.edu}
\affiliation{Associate, Nanoscale Device Characterization Division, National Institute of Standards and Technology, Gaithersburg, MD 20899, USA}
\affiliation{Department of Physics and Astronomy, West Virginia University, Morgantown, WV 26506-6315, USA}
\date{\today}

\begin{abstract}
Multi-exciton correlations shape the photo-induced response of nanostructured materials, particularly when interactions are enhanced by light confinement. Here multidimensional coherent spectroscopy is used to quantify biexciton and exciton-polariton dynamics in a semiconductor microcavity. One- and two-quantum spectra, which are dominated by polariton-related contributions, also include a polarization-dependent biexciton feature whose magnitude and spectral coordinates depend on the detuning between the cavity mode and the exciton resonance. Comparison of spectra acquired using collinear and noncollinear experimental geometries indicates that excitation wavevector plays no significant role in this behavior. The measured energy dispersion and cavity enhancement are not compatible with uncoupled biexcitons or bipolaritons. 
To explain the measurements, an indirect biexciton-photon coupling model is introduced whereby biexcitons are formed from dark excitons that Coulomb couple to the bright-exciton fraction of the polaritons. The model addresses inconsistent observations of biexcitons previously reported in semiconductor microcavities and is generalizable to any material where optically dark excitons contribute to light-matter interaction. Our results suggest a mechanism to access long-lived dark excitons through multi-exciton correlations in strongly coupled systems.
\end{abstract}

\maketitle

Optical traps that enhance the interaction of absorbers/emitters have fostered technological applications \cite{Schneider2013,Luo2023} and enabled the exploration of macroscopic coherent effects \cite{deng_condensation_2002} and exotic quantum states of matter \cite{Boulier2020,Trypogeorgos2025}. In planar microcavities containing semiconductor nanostructures \cite{weisbuch_observation_1992}, strong coupling hybridizes resonant cavity $\Ephoton$ and exciton $\EX$ modes into two or more exciton-polariton branches \cite{khitrova_nonlinear_1999,deng_exciton-polariton_2010,Toffoletti2025}. There has been an extended debate about the role of multi-exciton correlations in polariton many-body dynamics, particularly biexcitons (\emph{i.e.}~four-particle Coulomb correlations) in III-V semiconductor microcavities \cite{borri_biexcitons_2000,baars_biexciton_2001,borri_biexcitons_2003,Corfdir2012,Mitsumori2016,Choo2024}.
Time-resolved coherent spectroscopy has demonstrated that the polariton dynamics is modified by a Feshbach resonance \cite{takemura_polaritonic_2014} between the polariton and biexciton energies \cite{wilmer_multidimensional_2015, NavadehToupchi2019}. Experiments have also shown a transfer of oscillator strength from the saturated exciton transition to the biexciton \cite{saba_crossover_2000}, biexciton-mediated indirect exchange interactions \cite{Vladimirova2010}, and enhanced polariton cross-correlations \cite{wen_influence_2013, zhang_polarization-dependent_2007}. 
Despite extensive studies, however, it is not clear whether the biexciton couples directly to and is modified by the intracavity field.
Even for similar heterostructures in which biexcitons were readily observed, there is opposing evidence for their dependence on cavity-detuning $\Delta = \Ephoton-\EX$ \cite{baars_biexciton_2001,borri_biexcitons_2003,wilmer_multidimensional_2015}.

Understanding biexciton coupled states would offer a mechanism to harness the dynamics of multiexciton complexes, with potential applications in nanophotonics and quantum photon sources \cite{Oka2008,Carusotto_2010,Dousse2010,Cai2024}.
Direct probes of biexciton dispersion as a function of the detuning have been hampered by the relative weakness of biexciton transitions, spectral congestion, and limited resolution.
Multidimensional coherent spectroscopy (MDCS) can resolve many-body contributions to the photodynamics of a system \cite{ wen_influence_2013,wilmer_multidimensional_2015, Tollerud2016, Hao2016, Smallwood2018, Thouin2019,Muir2022, Li2022, Liang2022, Huang2023} by using a sequence of ultrashort pulses to disentangle excitation and emission transitions along separate frequency axes. However, due to their steep energy-momentum dispersion, polaritons are highly sensitive to the angle of the incident light, which influences their energy, scattering, and relaxation rates \cite{takemura_two-dimensional_2015}.
Biexciton contributions in a given experiment may therefore depend on the excitation geometry \cite{KuwataGonokami1997}.

Here, we resolve the coherent interaction between exciton-polaritons and biexcitons in a III-V semiconductor microcavity using both collinear (C-MDCS) and non-collinear (NC-MDCS) experimental geometries.
Biexcitonic features are identified and spectrally isolated using polarization-dependent selection rules.
Comparable detuning-dependent signatures obtained in both experimental geometries demonstrate that the biexciton's coherent response is essentially identical for excitation with multiple wavevectors.
Then we characterize the feature's magnitude and energy dispersion with detuning to understand the generation mechanism and the nature of its coupling to the cavity.
The biexciton's response is modified by the cavity, entailing an interaction with the polariton modes.
This observed relatively weak dispersion is incompatible with a bipolariton picture, which arises from direct coupling between the biexciton and the two-photon state \cite{borri_biexcitons_2000, Kudlis2025}.
Instead, it comes from an indirect coupling between the optically dark manifold of excitons localized by the fluctuation of the well potential \cite{glinka_coherent_2013} and strongly coupled bright excitons.

\begin{figure*}
\centering
\includegraphics[width=17.2cm]{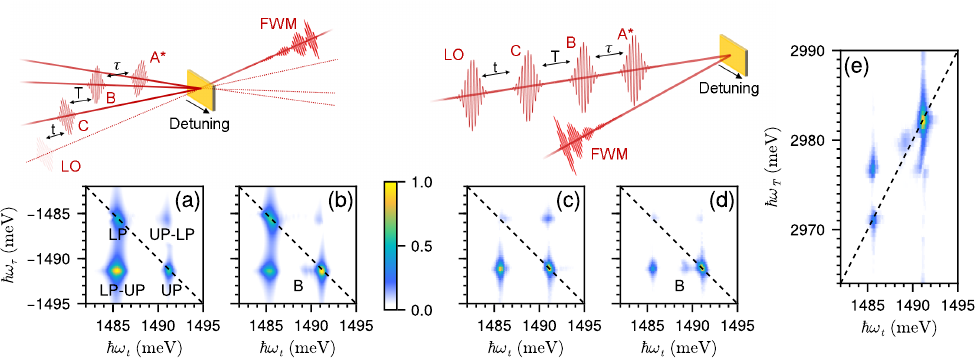}
\caption{Representative MDCS spectra showing the polarization-dependent biexciton feature. (a,b) NC-MDCS spectra for $\Delta = -4.9$~meV and $T=0.1$~ps (the inset shows the non-collinear geometry): (a) co-circular (++++) polarization and (b) co-linear (XXXX) polarization. (c,d) C-MDCS spectra for $\Delta = -4.0$~meV and $T=1$~ps (the inset shows the collinear geometry): (c) ++++ and (d) XXXX polarization recorded using NC-MDCS. (e) XXXX two-quantum spectrum measured using C-MDCS.}
\label{Fig: Figure_1}
\end{figure*}

The sample, grown by molecular beam epitaxy on a GaAs substrate, consists of two GaAs/AlAs distributed Bragg reflectors forming a wedged $\lambda$ cavity with a single 8 nm In$_{0.04}$Ga$_{0.96}$As quantum well (QW) at its center \cite{wilmer_multidimensional_2015}. Because of the wedge, $\Ephoton$ depends on the location of the excitation spot and also depends on the angle of incidence $\theta_i$ through the in-plane momentum $k_\parallel$.
For $\Ephoton \approx \EX$, the anticrossing between the lower polariton (LP) and upper polariton (UP) branches as a function of $\Delta$, within a Jaynes-Cummings approach, is given by
\begin{equation}\label{eq: polariton_dispersion}
    E_{\mathrm{UP},\mathrm{LP}} (\Delta)=\left[2\EX+\Delta\pm\sqrt{\Delta^2+\ORabi^2 }\right]/2,
\end{equation}
where $\ORabi$ is the vacuum Rabi splitting. We obtain $\EX \approx 1490.7$~meV and $\ORabi \approx 3.2$~meV for this sample at cryogenic temperature by using Eq.~(\ref{eq: polariton_dispersion}) to fit the energy of the polariton features in transmission or reflection spectra.

The NC-MDCS spectra were obtained by a Multidimensional Optical Nonlinear Spectrometer (MONSTR) \cite{bristow_versatile_2009,wahlstrand_automated_2019}, which is described in the Supplementary Material \cite{SM}.
In brief, three incident pulses with duration $\approx$ 120 fs and time delays $\tau$ and $T$ excite the sample, each impinging from a different direction but all with $\theta_i \approx 5^\circ$, as shown in the top left inset of Fig.~\ref{Fig: Figure_1}.
A fourth pulse acts as a local oscillator (LO) delayed by $t$ to measure the complex four-wave mixing (FWM) signal in transmission.
Time ordering is used to isolate rephasing one-quantum or non-rephasing two-quantum frequency correlation maps, which display absorption pathways involving singly excited (1Q) or doubly excited (2Q) manifolds respectively \cite{autry_excitation_2020}.
Figure \ref{Fig: Figure_1}ab shows the magnitude of one-quantum rephasing spectra for both co-circular polarization (++++, denoting the polarization of the excitation pulses and the FWM) and co-linear polarization (XXXX). The spectra shown were measured for $\Delta = -4.9$~meV at fixed $T=0.1$~ps.  Self- and cross-correlations of the UP and LP states appear as two diagonal peaks (LP and UP) and two cross peaks (LP-UP and UP-LP). An additional feature, labeled B, is only visible in the XXXX spectrum at an absorption energy $\hbar\omega_\tau = \EX$ and an emission energy $\hbar\omega_t = \EX - 1.3$~meV. This feature has been previously attributed to the biexciton \cite{wilmer_multidimensional_2015} because of its polarization dependence, which follows the spin selection rules for biexciton formation, and the observed offset between its absorption and emission energies, which is consistent with the binding energy $\Delta E_{b}^{\mathrm{B}}$ measured in bare GaAs quantum wells \cite{bristow_polarization_2009}.

In NC-MDCS, the nonlinear signal is spatially isolated from the linear background by using excitation pulses incident on the sample at different angles. As a result, each beam transfers a different in-plane momentum to the polariton, complicating the interpretation \cite{takemura_two-dimensional_2015} and possibly impacting on the observation of biexcitons, given their extended wave-function in momentum space \cite{Corfdir2012}.
To assess the influence of the wavevector difference on the spectra, we performed C-MDCS spectroscopy (see \cite{SM}) on the same sample.
In C-MDCS, the three excitation pulses as well as the LO pulse copropagate, and the FWM is measured in reflection, as shown in the right inset of Fig.~\ref{Fig: Figure_1}.
The resulting one-quantum rephasing spectra are shown in Fig.~\ref{Fig: Figure_1}cd for ++++ and XXXX polarizations for $\Delta \approx$ $-4.0$~meV and $T= 1$~ps. The observed difference in relative intensity of the LP and UP peaks in the C-MDCS and NC-MDCS may be caused by the different LO spectrum in the two cases, since in the C-MDCS setup the LO pulse interacts with the sample, while in NC-MDCS it passes around the sample.
In C-MDCS the biexcitonic feature B displays the same polarization dependence as in NC-MDCS, appearing for co-linear but not co-circular polarization. The C-MDCS spectra shown in Fig.~\ref{Fig: Figure_1}cd were obtained at $\theta_i = \mathrm{2.75^\circ}$. Measurements obtained for $0^\circ < \theta_i < 5^\circ$ \cite{SM} showed essentially the same spectrum, indicating that the B feature is not affected by the overall transferred in-plane momentum within this range.
Interestingly, a cross peak between the B and UP features is resolved in C-MDCS spectra, suggesting the presence of a relaxation process from the UP to a lower energy state near the bare exciton energy $\EX$.
Further evidence of the biexcitonic origin of the B feature is provided by the observation of an above-diagonal peak for XXXX polarization in double-quantum MDCS spectra \cite{karaiskaj_two-quantum_2010,wilmer_multidimensional_2016}, as shown in Fig.~\ref{Fig: Figure_1}e for C-MDCS.

\begin{figure}
\includegraphics[width=8.6cm]{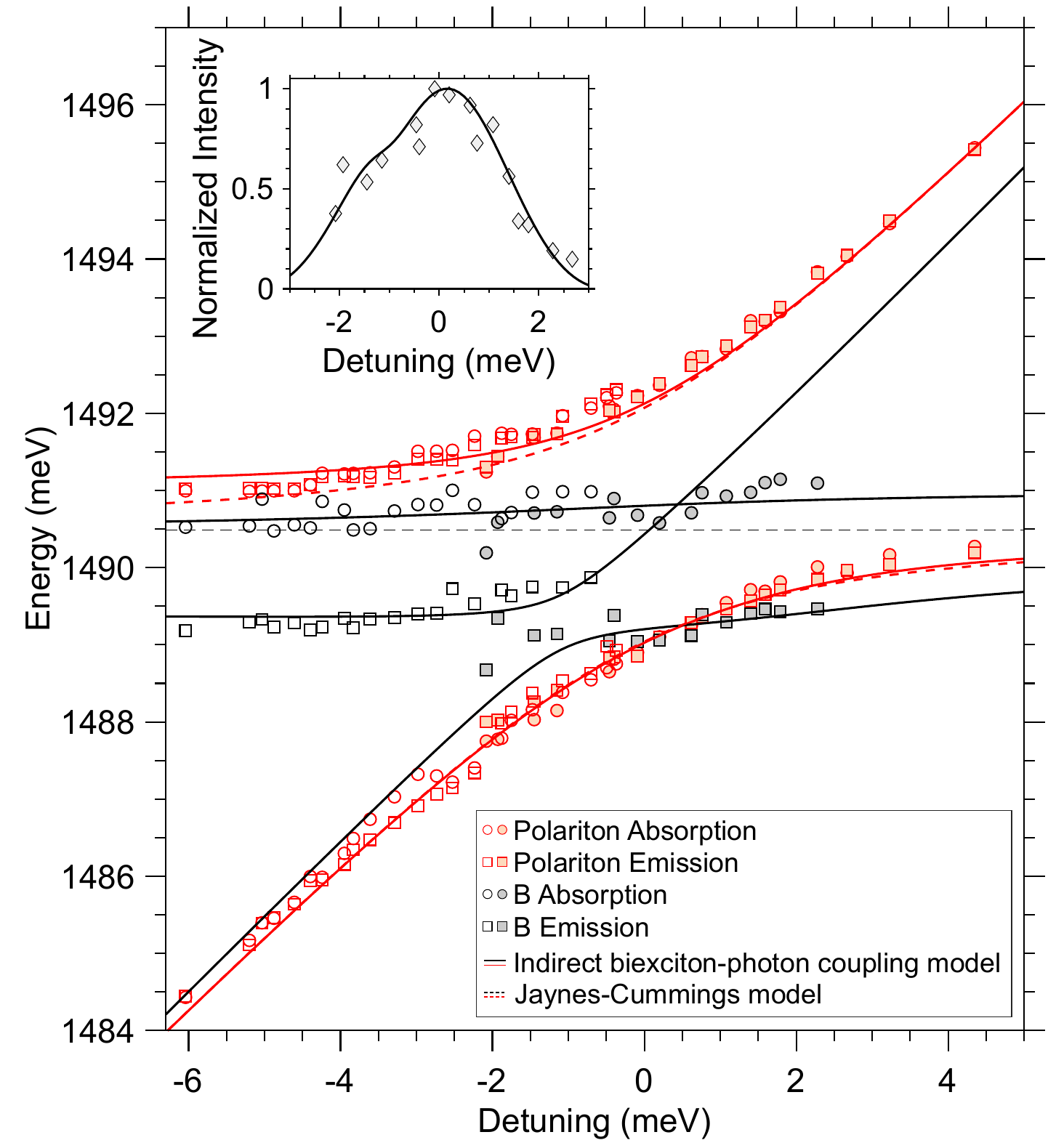}
\caption{Extracted absorption (circles) and emission (squares) energies from both NC-MDCS (filled) and C-MDCS (open) for the UP, LP, and B features as a function of $\Delta$. The inset shows the extracted magnitude of the B feature near $\Delta = 0$. Dashed lines show fits to Eq.~(1) and a fixed (dark) exciton energy, and solid lines show fits to the indirect biexciton-photon coupling model. Colors highlight branches related to the UP or LP (red) and to B (black).}
\label{Fig: Detuning_dependence}
\end{figure}

We next analyzed the detuning dependence of the spectral features, particularly for cross-circular polarization (-- -- ++) conditions, which exhibits the largest B peak intensity with respect to the polariton peaks. 
The B position and amplitude are extracted using linear prediction by singular value decomposition \cite{swagel_analysis_2021} or Voigt fitting of the antidiagonal lineslice through the peak (see \cite{SM} for details).
Figure \ref{Fig: Detuning_dependence} shows spectral coordinates of the B, UP and LP peaks along the absorption and emission energy axes and the extracted B magnitude (inset) as a function of $\Delta$.
The median uncertainty of the extracted spectral coordinates (95 \% confidence interval) is 0.08 meV; a full discussion of uncertainty is in \cite{SM}.
The $\Delta$-dependent absorption and emission energies of the LP and UP diagonal and cross peaks follow the polariton anticrossing dispersion. Their lineshapes and relative intensity are dominated by excitation induced effects (beyond the mean-field description) mediated by the $\Delta$-dependent exciton fraction \cite{nardin_coherent_2014,Smallwood2018}. The B peak shows a different behavior. Its absorption is close to the bare exciton energy, while its emission is red-shifted, as mentioned earlier. However, while the absorption and emission energies of B do not show the polariton dispersion, they are also not entirely independent of $\Delta$; they gradually increase over the measured range. The emission energy also shows a discontinuity around $\Delta \approx -1$~meV. Moreover, the B magnitude reaches a maximum at $\Delta\approx0$~meV, indicating a cavity enhancement.

Biexciton peaks in 1Q-rephasing MDCS spectra arise from excited-state absorption (ESA) pathways with coherences that mix 1Q and 2Q states during the time interval $t-T$, (\emph{e.g.}~$\ket{B}\bra{X}$; see \cite{SM} for details). The resulting ESA peak has an $\hbar \omega_t$ coordinate that reflects the energy difference between the states participating in such coherences. The detuning dependence therefore probes the extent to which the cavity affects the biexciton dynamics. If biexcitons form in the absence of any correlations with excitons coupled to the cavity, the B feature should be independent of $\Delta$ and relatively weak due to the lack of cavity enhancement. Biexcitons that instead directly couple to the two-photon states would form bipolaritons, which should exhibit a polariton-like dispersion with $\Delta$ and substantially overlap with the polariton features due to their expected small binding energy \cite{Borri2000}. Our observations exclude these two possibilities.
Instead, we model the B feature assuming that optically weak (\emph{dark}) states -- not directly coupled to the intracavity photon field -- affect the polariton dynamics due to Coulomb correlations with their bright exciton fractions.
This leads to a $\Delta$ dependence that is less pronounced than that of the polariton states, and also produces a cavity enhancement.

\begin{figure}
\includegraphics[width=8.6 cm]{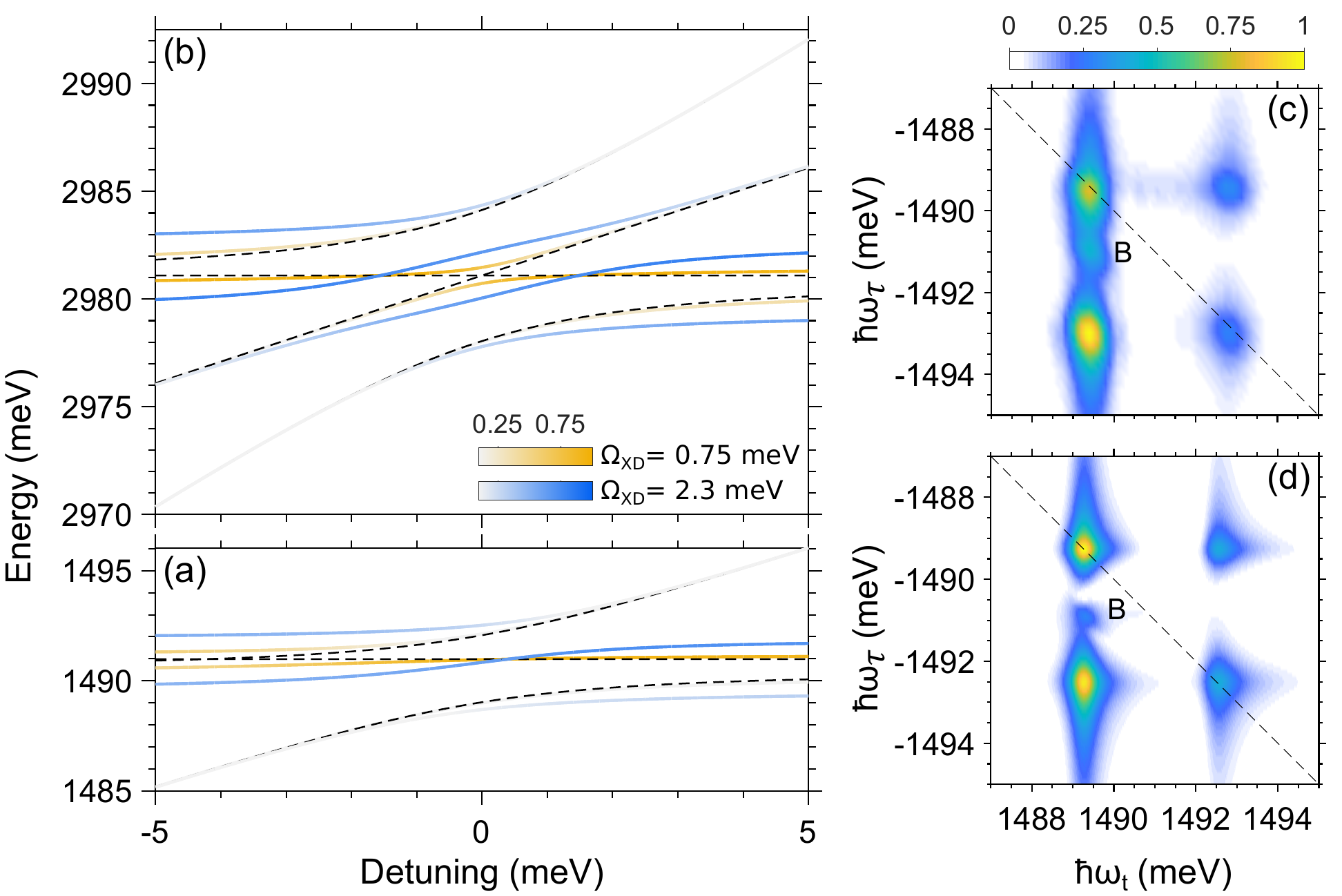}
\caption{Calculated polariton dispersion curves and MDCS spectra. Predicted energies of the states in the (a) 1Q and (b) 2Q manifolds for two values of the coupling $\Obiexc$. The color saturation indicates the magnitude of the Hopfield coefficient of the dark-exciton fraction of each state.  (c,d) NC-MCDS experimental (c)  and simulated (d) spectra at $\Delta=+0.62$~meV.}
\label{Fig: Fig3}
\end{figure}

We describe such dynamics in the exciton-photon basis with the Hamiltonian
\begin{equation}\label{eq: full_bx_H}
        H_0= H_{\mathrm{P}}+H_{\mathrm{D}}+H_{\mathrm{XD}},
\end{equation}
where
\begin{eqnarray} H_{\mathrm{P}} &=& \sum_{i} \left[\Ephoton a^{\dagger}_{C,i} a_{C,i} + E_{X}  \sigma_{X,i}^{\dagger}  \sigma_{X,i} \right. \nonumber \\ && + \left. \frac{1}{2}\ORabi \left( \sigma_{X,i}^{\dagger} a_{C,i}+a_{C,i}^\dagger \sigma_{X,i} \right) \right], \nonumber \\
        H_{\mathrm{D}} &=& \sum_{i} \Edark  \sigma_{D,i}^{\dagger}  \sigma_{D,i} + \sum_{i\neq j} \Delta E_{b}^{\mathrm{B}}\sigma^{\dagger}_{D,i}\sigma^{\dagger}_{D,j}\sigma_{D,i}\sigma_{D,j}, \nonumber \\ 
        H_{\mathrm{XD}} &=& \sum_i \frac{1}{2}\Obiexc \left( \sigma_{X,i}^{\dagger} \sigma_{D,i}+\sigma_{D,i}^{\dagger} \sigma_{X,i} \right), \nonumber
\end{eqnarray}
in which subscripts  $i\in \{+,-\}$ indicate the spin state.
Here $H_\mathrm{P}$ accounts for polariton formation through normal mode coupling ($\ORabi$) between bright excitons and the cavity, where  $\sigma_{X}$ and $a_C$ ($\sigma_{X}^{\dagger}$ and $a_C^{\dagger}$)  are the exciton and photon annihilation (creation) operators respectively. The first term in $H_{\mathrm{D}}$ accounts for a dark exciton with energy $\Edark$, while the second term represents the interaction of two correlated dark excitons with opposite spins, leading to the formation of a biexciton state in the 2Q manifold with binding energy $\Delta E_{b}^{\mathrm{B}}$. We hypothesize that such dark excitons originate from localized states within fluctuations of the QW potential \cite{albrecht_disorder_1996, rocca_biexcitons_1998,glinka_coherent_2013}.
The involvement of such states is supported by the large inhomogeneous broadening of the B feature along the diagonal direction observed in 1Q-rephasing spectra (Fig.~\ref{Fig: Figure_1}a-d).
This contrasts with the lineshape of the polariton features \cite{paul_coherent_2022}, which are known to be less sensitive to QW disorder \cite{Houdr1996,Whittaker1996,Wurdack2021}.
Finally, $H_{\mathrm{XD}}$ concerns the Coulomb interaction between bright and dark excitons through the coupling constant $\Obiexc$ and is responsible for the indirect coupling of biexcitons to the cavity.

Calculations using this model are shown in Fig.~\ref{Fig: Fig3}.
Eigenstates and eigenvalues of the system are calculated by diagonalizing the Hamiltonian with respect to an effective basis set (see \cite{SM} for details). In the 1Q manifold, there are three polariton branches that can be characterized by generalized Hopfield coefficients. If $\Obiexc=0$ (black dashed lines in Fig.~\ref{Fig: Fig3}a), the lower and upper branches correspond to polariton states with the usual dispersion, while the middle branch is the dispersionless dark-exciton state. If $\Obiexc\neq0$ (solid lines), the three branches hybridize and the middle branch becomes dispersive with $\Delta$. Similarly, there are four states in the 2Q manifold, as shown in Fig.~\ref{Fig: Fig3}b. The dark-exciton fraction (indicated by color saturation) and polariton fraction of each branch depends on the values of $\ORabi$, $\Obiexc$, $\Edark$, and $\Delta E_b^{\mathrm{B}}$. For our sample, the two middle 2Q branches carry the majority of the mixing, which involve the biexciton and mixed polariton states.  

We assume that the cavity-bright exciton subsystem interacts perturbatively with a classical external electromagnetic field through a quasi-mode coupling with the cavity photon \cite{takemura_two-dimensional_2015}, while the dark excitons participate through a dipole interaction. The interaction is given by
\begin{equation}\label{eq: Hint}
H'=  \left(\mu_{qm}^* a_C^\dagger+{\mu}^*_{D} \sigma^{\dagger}_{D,i} \right)\E(t) + h.c.,
\end{equation}
where $\E(t)$ is the positive frequency component of the optical field,  $\mu_{qm}$ is the quasi-mode coupling strength and $\mu_{D}$ is the transition dipole moment of the dark excitons. Since the latter is small, the dark exciton can be anharmonically boosted by an indirect coupling with the polaritons \cite{Lttgens2022}, \emph{i.e.} $\mu_{D}^{1Q}<< \mu_{D}^{2Q}$, enhancing the B peak in MDCS spectra. We also accounted for the background interaction of polaritons and the reservoir of long-lived dark-exciton states by including a mean-field correction to the branch energies \cite{Takemura_PRB2014,Takemura_PRB2017} and for excitation-induced effects \cite{nardin_coherent_2014}, as detailed in \cite{SM}.
Within this formalism, we calculated MDCS spectra using the nonlinear response theory \cite{abramavicius_coherent_2009, yang_dissecting_2008,Yang2008} by means of a third-order expansion of the time-dependent perturbation $H'$ and a sum over the diagonal states of $H_0$ (see \cite{SM} for the full derivation). With the additional inclusion of polariton population relaxation at fixed detuning \cite{SM}, the calculated spectrum qualitatively reproduces the experimental spectrum, as shown in Fig.~\ref{Fig: Fig3}cd  for $\Delta=0.6$~meV.

Extracted peak energies and B peak amplitude for a best fit of the parameters of Eqs.~(\ref{eq: full_bx_H},\ref{eq: Hint}) are shown as solid lines in Fig.~\ref{Fig: Detuning_dependence}, obtained for a coupling $\Obiexc=0.75$~meV. We did not include polariton relaxation effects in the fit, since they mainly impact on the polariton peaks and highly depend on the details of the relaxation process \cite{Mondal2023}. The absorption energy of the B peak follows the $\Delta$ dependence of the hybrid middle branch in the 1Q manifold. The emission energy is well fit by the difference between this 1Q state and either the lower or the upper of the 2Q middle branches. The B peak magnitude dependence on $\Delta$ (inset of Fig.~\ref{Fig: Detuning_dependence}) is also well reproduced by the model. We note that contributions from the two middle branches peak at different $\Delta$ due to their different dark exciton and cavity fractions. For example, for $\Delta>-1$~meV, only the lower branch contributes significantly to the spectrum, while we only observe the upper branch for $\Delta<-2.5$~meV. Additional modifications of their relative weight in the spectra might be related to a fast relaxation between the two branches that is not resolvable within the duration of our probe pulses.

In summary, we observed a biexcitonic feature in C- and NC-MDCS that appears to be independent of the experimental geometry, but does however show cavity enhancement and an energy shift as a function of detuning.
We introduce an indirect biexciton-photon coupling mechanism that explains these observations and may help resolve the opposing evidence of the biexciton's detuning dependence in previous literature.
In our model, biexcitons are formed from dark excitons which Coulomb couple to the bright-exciton fraction of the polaritons.
Depending on the coupling strength, the biexcitonic feature may appear for difference ranges of detuning, which justifies why the cavity affects the biexciton differently in similar heterostructures. 

Our results show that the biexciton's detuning dependence can probe coupling between the dark and bright excitons, and that an indirect biexciton-photon coupling mechanism underpins how  dark excitons influence the coherent response of multi-exciton complexes.
This may be of particular interest in transition metal dichalcogenides and layered perovskites, where it has been shown that dark excitons contribute pivotally to the polariton dynamics \cite{Wei2023, Fitzgerald_2024,Ferreira_2024,Fieramosca_2024}. Access to long-lived dark states via the bright states may allow for suitable coherent control protocols to be developed \cite{Ma2022, Chang2024} and impact on the optimization strategies of non-classical light sources for quantum technology applications \cite{Dousse2010,Lehner2023}.

\section{Acknowledgments}
\begin{acknowledgments}
The authors acknowledge assistance from E.~Swagel and thank M. Munoz, S. Pookpanratana, and V. Aksyuk for critical readings of the manuscript.
\end{acknowledgments}

\input{main.bbl}

\end{document}


\title{Supplementary Material: Biexciton-polariton coupling mediated by dark states}
\author{G.~Fumero}
\email{giuseppe.fumero@gmail.com}
\affiliation{Associate, Nanoscale Device Characterization Division, National Institute of Standards and Technology, Gaithersburg, MD 20899, USA}
\affiliation{Department of Physics and Astronomy, West Virginia University, Morgantown, WV 26506, USA}
\author{J.~Paul}
\affiliation{Associate, Nanoscale Device Characterization Division, National Institute of Standards and Technology, Gaithersburg, MD 20899, USA}
\affiliation{Department of Physics and Astronomy, West Virginia University, Morgantown, WV 26506, USA}
\author{J.~K.~Wahlstrand}
\affiliation{Nanoscale Device Characterization Division, National Institute of Standards and Technology, Gaithersburg, MD 20899, USA}
\author{A.~D.~Bristow}
\email{alan.bristow@mail.wvu.edu}
\affiliation{Associate, Nanoscale Device Characterization Division, National Institute of Standards and Technology, Gaithersburg, MD 20899, USA}
\affiliation{Department of Physics and Astronomy, West Virginia University, Morgantown, WV 26506, USA}
\date{\today}

\maketitle
\section{Experimental details}
The non-collinear multidimensional coherent spectroscopy (NC-MDCS) setup uses a Ti:sapphire laser outputting 120 fs pulses at 76 MHz. Three excitation pulses are generated by a Multidimensional Optical Nonlinear Spectrometer (MONSTR) \cite{bristow_versatile_2009}, which outputs them in a box geometry (A B, C in the left inset of Fig.~1 of the main text). The pulses are then focused using a common 20 cm lens, such that they each have an incident angle of $\theta_i \approx 5^\circ$ with respect to the normal axis of the sample, corresponding to in-plane wavevector $k_{\parallel}\approx 0.68$ cm$^{-1}$.  The four-wave mixing (FWM) signal along the phase-matched direction defined by $\bk_{\mathrm{FWM}}=-\bk_{\mathrm{A}}+\bk_{\mathrm{B}}+\bk_{\mathrm{C}}$ is measured interferometrically with respect to a reference pulse bypassing the sample, which was kept at $\approx 12$~K. Three translation stages control time delays between the first and second (second and third) pairs of pulses, defined as $\tau$ ($T$). The FWM emission time coordinate $t$ is defined with respect to the C pulse. Time ordering is used to isolate the rephasing one-quantum, $S_{1Q} (-\omega_\tau,T,\omega_t)$, or non-rephasing two-quantum, $S_{2Q} (\tau, \omega_T,\omega_t)$, frequency correlation maps displaying absorption pathways that involve one-particle or two-particle states, respectively. One spectral axis, $\omega_\tau$ or $\omega_T$, is determined by a numerical Fourier transform while $\omega_t$ is obtained by spectral interferometry. The spectral interferograms are detected using a CCD camera and a high-resolution spectrometer. The polarization of each pulse is separately controlled with variable retarders \cite{wahlstrand_automated_2019} and phase cycling \cite{munoz_fast_2020} is used to improve the signal-to-noise ratio. All NC-MDCS spectra are recorded with a total excitation fluence of
approximately $50$~nJ/cm$^2$ and a $1/e^2$ spot diameter of approximately 15~$\mu$m, equivalent to an excitation density of
$10^{11}$~cm$^{-2}$.

The collinear multidimensional coherent spectroscopy (C-MDCS) setup uses a Ti:sapphire laser outputting 15 fs pulses with a 79 MHz repetition rate. A $4f$ pulse shaper was used to narrow the bandwidth of the laser to the wavelength range near the exciton energy. Isolation of the nonlinear signal is achieved by using radio-frequency modulation of the excitation beams A, B, and C, and the local oscillator pulse D (which interacts with the sample) and demodulation through a lock-in amplifier \cite{nardin_multidimensional_2013}, using a commercial spectrometer. After the spectrometer, the train of four pulses is focused by a 5 cm lens on the sample, which was held at approximately 8~K, and the signal is detected in a reflection geometry. Polarization is controlled using static $\lambda/2$ waveplates placed inside the spectrometer and a $\lambda/4$ waveplate placed just before the focusing lens. The total excitation fluence is varied between 40~nJ/cm$^2$ and 110~nJ/cm$^2$.

\section{Theoretical model}
\subsection{Nonlinear response theory for multidimensional coherent spectroscopy}
\begin{figure}[ht]
    \centering
    \includegraphics[width=0.8\linewidth]{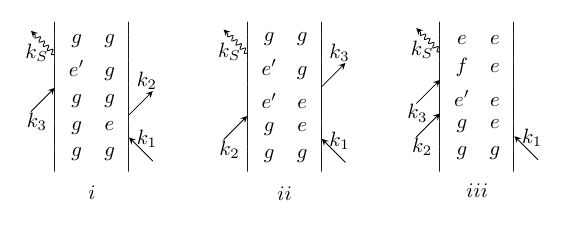}
    \caption{Feynman diagrams corresponding to the GSB (i), ESE (ii) and ESA (iii) pathways contributing to the 1Q-rephasing MDCS spectra. }
    \label{fig:diagrams}
\end{figure}
We briefly summarize here the derivation of the spectroscopic signal for multidimensional coherent spectroscopy (MDCS) using the nonlinear response theory \cite{yang_dissecting_2008}. In this formalism, the nonlinear polarization $\bm{P}^{(n)}$ generated in the sample by interaction with three delayed pulses is calculated perturbatively, assuming a full Hamiltonian
\begin{equation}
    H=H_0+H',
\end{equation}
where $H_0$ describes the free evolution of the system and $H'$ is a perturbative term accounting for the interaction with the light. The total electric field of the excitation pulses driving the interaction is given by
\begin{equation}
    \boldsymbol E(\bm r,t)=\sum_{n=1,2,3}\boldsymbol\epsilon_n \bar\E_n(t-t_n) e^{-i \omega_n (t-t_n) +i\boldsymbol{k_n \cdot r}}+\mathrm{c.c.},
\end{equation}
where $\boldsymbol \epsilon_n$, $\bar\E_n$, $t_n$, $\omega_n$ and $\boldsymbol{k_n}$  are the polarization unit vector, temporal envelope,  delay time, carrier frequency and wavevector of the $n^\mathrm{th}$ pulse. We consider a dipole interaction
\begin{equation}\label{eq: dipoleH}
    H'=- \bm V \cdot \bm E(\bm r,t),
\end{equation}
where $\bm V=  \bm{\mu} + \bm{\mu}^\dagger$ is the dipole operator. The source of the MDCS signal is the third-order component of the nonlinear polarization $\bm P^{(3)}=\mathrm{Tr}\left(\bm V\rho^{(3)} \right)$, which is calculated by computing the expectation value of the dipole operator using iterative integration of the Liouville equation for the system's density matrix $\rho$ up to the third-order in $H'$: 
\begin{equation}
\begin{split}
    \bm P^{(3)}(\bm r, t)&=\mean{\bm V }=\mathrm{Tr}(\bm V \rho^{(3)}(\bm r,t))=\\
    &=\int^{t}_{-\infty} dt_3 \int^{t_3}_{-\infty} dt_2 \int^{t_2}_{-\infty} dt_1 R(t,t_1,t_2,t_3) \bm E(\bm r,t_3) \bm E(\bm r,t_2) \bm E(\bm r,t_1).
\end{split}
\end{equation}
We isolate the 1Q-rephasing component of the MDCS signal imposing the phase matching condition $k_S=-k_1+k_2+k_3$ and spatially integrating the polarization over the interaction volume. The response function $R(t,t_1,t_2,t_3)$ can be computed using the Feynman diagrams in Fig.~\ref{fig:diagrams},
\begin{equation}
    \begin{split}
        &R(t,t_1,t_2,t_3)=\left(\frac{i}{\hbar}\right)^3\left[R_i(t,t_1,t_2,t_3)+R_{ii}(t,t_1,t_2,t_3)+R_{iii}(t,t_1,t_2,t_3)\right],\\
        &R_{i}(t.t_1,t_2,t_3)=\mathrm{Tr}\left(\muv(t)\muv^\dagger(t_3)\rho(-\infty)\muv^\dagger (t_2) \muv(t_1) \right),\\
        &R_{ii}(t.t_1,t_2,t_3)=\mathrm{Tr}\left(\muv(t)\muv^\dagger(t_2)\rho(-\infty)\muv^\dagger (t_3) \muv(t_1)\right), \\
        &R_{iii}(t.t_1,t_2,t_3)=-\mathrm{Tr}\left(\muv(t)\muv^\dagger(t_3)\muv\dagger(t_2)\rho(-\infty) \muv(t_1) \right),\\
    \end{split}
    \label{eq: response}
\end{equation}
where $\muv(t')=G^\dagger(t')\muv G(t')$ is the dipole operator in the interaction picture and $G(t-t')=\theta(t-t') e^{-\frac{i}{\hbar}H_0 (t-t')}$ is the Green function of the unperturbed Hamiltonian $H_0$. The three terms in Eq.~(\ref{eq: response}) account for ground-state bleach (GSB) and excited-state emission (ESE) from the 1Q manifold of $H_0$, as well as the excited-state absorption (ESA) involving both the 1Q and 2Q manifolds. In the impulsive limit where the field envelopes are much shorter than the delay periods and the time scale of the system dynamics, we can treat the temporal envelopes of the excitation field as delta functions that are used to solve the integrals,
\begin{equation}
    \bar\E_n(t)  =\E_n \delta(t-T_n).
\end{equation}
The amplitude of the spectroscopic signal $S^{(3)}(\omega)$ is given by the interferometric transmission of the nonlinear field $\E_s(t) \propto i P^{(3)}(t)$ with a delayed fourth pulse $E_{LO}(t)=\E_{LO}e^{-i\omega (t-T_4)}+\mathrm{c.c.}$, which acts as local oscillator,
\begin{equation}
    S^{(3)}(\omega)=\mathcal{I}m \int_{-\infty}^{+\infty}dt \,\E_{LO}(\omega)e^{i \omega t} P^{(3)}( t).
\end{equation}
We define time delays $t_{abs}= T_2-T_1$, $T=T_3-T_2$ and $t_{em}=T_4-T_3$.
Note that the time delays $\tau$ and $t$ defined in the main text are the same as $t_{abs}$ and $t_{em}$, respectively.
We consider a system with one ground state $g$ and two excited-state manifolds: one composed by singly exited states $\{e,e',\dots\}$ (1Q manifold) and the other by doubly excited states $\{f,f',\dots\}$ (2Q manifold). Expanding the Green functions over the eigenstates of $H_0$ and performing a two-dimensional Fourier transform of the delay $t_{em}$ and $t_{abs}$, we obtain the sum-over-states (SoS) expression of the third-order signal:
\begin{subequations}\label{eq: SoS}
    \begin{align}\label{eq: SoS-GSB}
        &S^{(3)}_{i}(\Omega_{em},T, \Omega_{abs})\propto +\sum_{e,e'} \mathcal{I}m \frac{\mu_{eg}(\hat{\epsilon}_1)\mu_{ge}(\hat{\epsilon}_2)\mu_{e'g}(\hat{\epsilon}_3)\mu_{ge'}(\hat{\epsilon}_4)\E_{LO}^*\E_3\E_2\E_1^*}{(\Omega_{em}-\omega_{e'g}+i\gamma_{e'g})(\Omega_{abs}+\omega_{eg}+i\gamma_{eg})},\\ \label{eq: SoS-SE}
        &S^{(3)}_{ii}(\Omega_{em},T, \Omega_{abs})\propto +\sum_{e,e'}\mathcal{I}m \frac{\mu_{eg}(\hat{\epsilon}_1)\mu_{e'g}(\hat{\epsilon}_2)\mu_{ge}(\hat{\epsilon}_3)\mu_{ge'}(\hat{\epsilon}_4) e^{-i(\omega_{e'e}-i\gamma_{e'e})T}\E_{LO}^*\E_3\E_2\E_1^*}{(\Omega_{em}-\omega_{e'g}+i\gamma_{e'g})(\Omega_{abs}+\omega_{eg}+i\gamma_{eg})},\\ \label{eq: SoS-ESA}
        &S^{(3)}_{iii}(\Omega_{em},T, \Omega_{abs})\propto -\sum_{e,e',f}\mathcal{I}m\frac{\mu_{eg}(\hat{\epsilon}_1)\mu_{e'g}(\hat{\epsilon}_2)\mu_{fe'}(\hat{\epsilon}_3)\mu_{ef}(\hat{\epsilon}_4) e^{-i(\omega_{e'e}-i\gamma_{e'e})T}\E_{LO}^*\E_3\E_2\E_1^*}{(\Omega_{em}-\omega_{fe}+i\gamma_{fe})(\Omega_{abs}+\omega_{eg}+i\gamma_{eg})},
    \end{align}
\end{subequations}
where $\omega_{ij}=\omega_{i}-\omega_{j}$ and $\gamma_{ij}=\frac{1}{2}\left(\gamma_i+\gamma_j\right)$ are the energy differences and homogeneous dephasing rates of eigenstates $i$ and $j$, while $\mu_{ij}(\epsilon_k)$ is the scalar product between the dipole element $\bra{i}\boldsymbol{\mu}\ket{j}$ and the polarization unit vector of the $k$-th excitation field. We note that using this approach, the spectral position and amplitude of the B peak as a function of $\Delta$ can be easily extracted by isolating the relevant terms in the SoS expansion. In particular, the first term in the denominator of the ESA contribution $S^{(3)}_{iii}$ stems from pathways involving the excitation of coherences $\omega_{fe}$ that mix 1Q and 2Q states. An example of such a pathway involving the B feature is
\begin{equation}
    \underbrace{\ket{g}\bra{g}}_{\mathrm{Initial\,state\,at\,} t_0} \rightarrow \underbrace{\ket{0}\bra{X}}_{\tau-t_0} \rightarrow \underbrace{\ket{B}\bra{X}}_{T-\tau} \rightarrow \underbrace{\ket{B}\bra{X}}_{t-T}\rightarrow \underbrace{\ket{X}\bra{X}}_{\mathrm{Final\,state\,at\,} t},
\end{equation}
where the brackets show the evolution of the density matrix  from the initial state at time $t_0$ to the final state at time $t$ and during  the three time intervals. Here $\ket{X}$ and $\ket{B}$ are the exciton and biexciton states respectively.
\subsection{Excitonic model}
In order to simulate the experimental MDCS spectra from Eqs.~(\ref{eq: SoS}), the dipole elements $\mu_{ij}$ and the eigenvalues $\om_{ij}=\omega_{ij}-\gamma_{ij}$ need to be calculated from a model unperturbed Hamiltonian $H_0$. Assuming that the MDCS pulses are weak and short compared to the time scale of few-particle scattering events, we model the interaction between dark excitons and polaritons as a quasi-equilibrium process, in which the density of the bright and dark states remains constant during the light-matter interactions. The unperturbed Hamiltonian is then described by $H_0$ in Eq.~(2) in the main text, which can be separately block-diagonalized in each manifold, in which the number of particles is fixed. We can now define an effective basis set for the 1Q and 2Q manifolds and diagonalize the relative block Hamiltonian to obtain the energies and dephasing of the states and the transition matrix elements from the eigenvalues and the generalized Hopfield coefficients, then use these to compute the SoS expressions of the MDCS signal. We consider a three-state  basis set spanning the 1Q manifold, $\{\kket{1}{0}\ket{0}, \kket{0}{1}\ket{0},\kket{0}{0}\ket{1}\}$, where the notation $\kket{i}{j}\ket{k}$ indicates a state with $i$ bright excitons, $j$ photons and $k$ dark excitons. From the basis set, we obtain an effective Hamiltonian
\begin{equation}\label{eq: 1Qbiexc}
    H^{(1Q)}=\begin{pmatrix}
\EX &  \frac{1}{2} \ORabi &  \frac{1}{2} \Obiexc \\
\frac{1}{2} \ORabi & E_{C} & 0\\
\frac{1}{2} \Obiexc & 0 & E_{D}
\end{pmatrix}.
\end{equation}
We note that the 1Q block is equivalent to the effective Hamiltonian used by Takemura \emph{et al.}~to model the Feshbach interaction between polariton states and the biexciton \cite{takemura_polaritonic_2014}.
Diagonalization leads to three branches. If the biexciton Coulomb coupling is zero, $\Obiexc=0$, the lower and upper branches correspond to the polariton states, with the regular detuning dispersion, while the middle branch is the dispersionless dark-exciton state. If $\Obiexc\neq0$, the three branches hybridize, with fractions depending on the coupling strength and the relative energy mismatches. Consequently, the middle branch becomes dispersive as a function of $\Delta$.
Similarly, the 2Q effective Hamiltonian is constructed from the basis set
\begin{math}
    \{\kket{2}{0}\ket{0}, \kket{1}{1}\ket{0}, \kket{0}{2}\ket{0}, \kket{0}{0}\ket{2}\},
\end{math} 
yielding
\begin{equation}\label{eq: 2Qbiexc}
    H^{(2Q)}=\begin{pmatrix}
2 \EX &  \frac{\sqrt{2}}{2} \ORabi & 0 & \frac{\sqrt{2}}{2} \Obiexc \\
\frac{\sqrt{2}}{2} \ORabi & \EX+E_{C} & \frac{\sqrt{2}}{2} \ORabi & 0\\
0 & \frac{\sqrt{2}}{2} \ORabi & 2 E_{C} & 0\\
\frac{\sqrt{2}}{2} \Obiexc & 0 & 0 & 2 E_{D}-\Delta E^{B}_{b}
\end{pmatrix}.
\end{equation}
Diagonalizing the 2Q Hamiltonian leads to four states. Depending on the biexciton and Rabi coupling strengths, the exciton energy and biexciton binding energy, the mixing between dark states and polaritons will impact differently on different branches.

The transition dipole elements are calculated by sandwiching $H'$ (defined in the main text) with the eigenstates of $H_0$. We can then define generalized Hopfield coefficients $\xi^{(1Q)}({p,\zeta})$ and $\xi^{(2Q)}({p,\zeta})$, obtained by the basis transform matrix of the diagonalization of Eqs.~(\ref{eq: 1Qbiexc},\ref{eq: 2Qbiexc}). The index $\zeta$ refers to the exciton basis, while $p$ refers to the polaritonic basis. We assign a $\zeta$ index to each of the basis states, as shown in Table \ref{tbl: zeta_index}.
\begin{table}[ht]
\caption{\label{tbl: zeta_index}
Assignment of the index $\zeta$ to the states of the polariton basis in the 1Q and 2Q manifolds.}
\centering
\begin{tabular}{llll|llll}
\multicolumn{4}{c|}{\cellcolor[HTML]{C0C0C0}1Q}                                  & \multicolumn{4}{c}{\cellcolor[HTML]{C0C0C0}2Q}                               \\ \hline
\multicolumn{1}{l|}{$\zeta$}     & \multicolumn{1}{l|}{1} & \multicolumn{1}{l|}{2} & 3 & \multicolumn{1}{l|}{1} & \multicolumn{1}{l|}{2} & \multicolumn{1}{l|}{3} & 4 \\ \hline
\multicolumn{1}{l|}{State} & \multicolumn{1}{l|}{$\kket{1}{0}\ket{0}$} & \multicolumn{1}{l|}{$\kket{0}{1}\ket{0}$} & $\kket{0}{0}\ket{1}$  & \multicolumn{1}{l|}{$\kket{2}{0}\ket{0}$} & \multicolumn{1}{l|} {$\kket{1}{1}\ket{0}$} & \multicolumn{1}{l|}{$\kket{0}{2}\ket{0}$} & $\kket{0}{0}\ket{2}$
\end{tabular}
\end{table}
Using this notation, we obtain the following transition matrix elements:
\begin{equation}
    \begin{split}
        &\mu_{ge_l}= \mu_{qm}\xi^{(1Q)}(l,2)+\mu_{D}\xi^{(1Q)}(l,3) \quad \mathrm{with}\, l=1,2,3\\
        &\mu_{gf_k}=0 \quad \mathrm{with}\, k=1,\dots,4\\
        &\mu_{e_1f_k}= \sqrt{2}\xi^{(1Q)}(1,2)\mu_{qm}^{(2Q)}\xi^{(2Q)}(k,3)+\xi^{(1Q)}(1,1)\mu_{qm}^{(2Q)}\xi^{(2Q)}(k,2) \quad \mathrm{with}\, k=1,\dots,4\\
        &\mu_{e_2f_k}= \sqrt{2}\xi^{(1Q)}(2,3)\mu^{(2Q)}_{D}\xi^{(2Q)}(k,4)+\xi^{(1Q)}(2,2)\mu_{qm}^{(1Q)}\xi^{(2Q)}(k,2) \quad \mathrm{with}\, k=1,\dots,4\\
        &\mu_{e_3f_k}= \sqrt{2}\xi^{(1Q)}(3,2)\mu_{qm}^{(2Q)}\xi^{(2Q)}(k,3)+\xi^{(1Q)}(3,1)\mu_{qm}^{(2Q)}\xi^{(2Q)}(k,2) \quad \mathrm{with}\, k=1,\dots,4\\
    \end{split}
\end{equation}
We note that, due to the coupling constants $\ORabi$ and $\Obiexc$ there are additional contributions to the transition matrix elements with respect to the Jaynes-Cummings model, coming from the cavity and biexciton fractions of all the states and mediated by $\mu_{qm}$ and $\mu_{D}$, respectively. 

\subsection{Background interaction with reservoir excitons}
For $\Delta> 2$~meV ($\Delta< 2$~meV), the calculated LP (UP) energies in the 1Q manifold blue-shift (red-shift)  from the values of the non-interacting model, due to the coupling to the dark exciton. This shift can be partially compensated by phenomenologically including polariton scattering with incoherent dark- and bright-exciton reservoirs by a mean-field treatment \cite{Takemura_PRB2014,Takemura_PRB2017}. For scattering of the polariton with the dark-exciton reservoir, the correction to the three polariton branches $E^{1Q}_k$ in the 1Q manifold is given by 
\begin{equation}\label{eq: backgroundInt}
    \Delta E^{1Q}_k = \\ \left(g_{\mathrm{spin}} -g_{B}^2\frac{E_{B}-2 E^{1Q}_k}{\left(E_{B}-2 E^{1Q}_k \right)^2+ \gamma_{B}^2} \right)|\xi^{(1Q)}(p,2)|^2,
\end{equation}
where $g_{\mathrm{spin}}$ and $g_{B}$ are the spin and biexciton interaction constants, $\xi^{(1Q)}(p,2)$ is the dark-exciton fraction of the polariton branch $p$, $E_B= 2 E_{D}-\Delta E^{B}_{b}$ and $\gamma_{B}$ is the biexciton broadening. Here we considered exciton-polariton interactions as opposed to polariton-polariton interactions studied in \cite{Takemura_PRB2014}, so the correction scales with the square of the excitonic fraction instead of the fourth power. Although not necessary to explain the $\Delta$ dependence of the B peak, the existence of a long-lived reservoir is compatible with our observation of a non-vanishing polariton signal at long $T$ delays and with recent observations in similar samples \cite{Schmidt2019,Rozas2023}. 
\subsection{Homogeneous and inhomogeneous broadening}
Homogeneous broadening is directly included in Eqs.~(\ref{eq: SoS}) by converting the dephasing rates $\gamma$ from the exciton to the polariton basis using the generalized Hopfield coefficients $\xi^{(1Q)}({p,\zeta})$ and  $\xi^{(2Q)}({p,\zeta})$. For the three branches in the 1Q manifold, we have
\begin{equation}
\gamma^{(1Q)}_p= |\xi(p,1)|^2 \gamma_{X}+|\xi(p,2)|^2 \gamma_{D}+|\xi(p,3)|^2 \gamma_{C} \quad \mathrm{with}\, p=1,\dots,3,
\end{equation}
where $\gamma_{X}$, $\gamma_{D}$, and $\gamma_{C}$ are the dephasing rates of the bright exciton, the dark exciton, and the cavity mode, respectively. For the four polaritonic branches in the 2Q manifold, we obtain
\begin{equation}
\gamma^{(2Q)}_p= |\xi(p,1)|^2 \gamma_{X}+\frac{1}{2}|\xi(p,2)|^2 (\gamma_{X}+\gamma_{c})+|\xi(p,3)|^2 \gamma_{C}+|\xi(p,4)|^2 \gamma_{B} \quad \mathrm{with}\, p=1,\dots,4.
\end{equation}
Inhomogeneous broadening is added by convolving the lineshapes obtained by Eqs.~(\ref{eq: SoS}) with a Gaussian lineshape function,
\begin{equation}
    S^{(3)}(\Omega_{em},T, \Omega_{abs})=\sum_{n=1}^3\int d\omega' e^{-2\log 2 \left(\frac{\omega'}{b_D}\right)^2} S_n(\Omega_{em},T, \Omega_{abs}; \EX+\frac{b_{X}}{b_D}\omega',\Edark+\omega'),
\end{equation}
where $b_D$ and $b_{X}$ are the inhomogeneous broadenings of the dark and bright excitons and $S_n$ are the SoS expressions of Eqs.~(\ref{eq: SoS}), calculated using $\EX+\frac{b_{X}}{b_D}\omega'$ and $\Edark+\omega'$ as bright- and dark-exciton energy, respectively.
\subsection{Excitation-induced effects}
To further account for many-body interactions of excitons, excitation-induced shift (EIS) and excitation-induced dephasing (EID) are phenomenologically included in the model by adding a correction to the energy and dephasing rates of the 2Q states \cite{nardin_coherent_2014}. The correction for each branch is weighted by the corresponding excitonic fraction through the generalized Hopfield coefficient $\xi(p,1)$,
\begin{equation}
\begin{split}
    \gamma^{(2Q,EID)}_p&= \gamma^{(2Q)}_p+ |\xi(p,1)|^2 \Delta \gamma^{(EID)},\\
    E^{(2Q,EIS)}_{p}&=E^{(2Q)}_{p}+|\xi(p,1)|^2\Delta E^{\mathrm{(EIS)}},
\end{split}
\end{equation}
where $p=1,\dots,4$ and $E^{(2Q)}_{p}$ are the eigenvalues of the 2Q Hamiltonian in Eq.~(\ref{eq: 2Qbiexc}).
\subsection{Population relaxation}
We can partially account for population transfer between the polaritonic branches during the delay $T$ by adopting a simplified secular Redfield theory of  relaxation, as discussed in \cite{abramavicius_coherent_2009}. In this framework, relaxation of populations is decoupled from that of coherences so that the former can be described by the Pauli Master equation:
\begin{equation}\label{eq: Pauli}
    \frac{d}{dt}\rho_{ee}(t)=-\sum_{e'} K_{e\,e,e'e'}\rho_{e'e'}(t),
\end{equation}
where $K_{e\,e,e'e'}$ are the population transfer rates from level $e'$ to $e$.
Coherences satisfy the Liouville equation
\begin{equation}\label{eq: Liouville}
    \frac{d}{dt}\rho_{e'e}(t)=\left[-i\omega_{e'e}-\frac{1}{2}(K_{e\,e,e\,e}+K_{e'e',e'e'})-\gamma_{e'e}\right]\rho_{e'e} (t),
\end{equation}
where the dephasing includes both the pure dephasing term $\gamma_{e'e}$, discussed in the previous sections, and the population relaxation $K_{i\,i,i\,i}$. In our case, since during $T$ the system evolves in the 1Q manifold, $e$ and $e'$ are eigenstates of Eq.~(\ref{eq: 1Qbiexc}) and $K$ is a $3\times 3$ matrix,
\begin{equation}\label{eq: K}
    K=\begin{pmatrix}
K_{1\,1,1\,1} & K_{1\,1,2\,2} & K_{1\,1,3\,3} \\
K_{2\,2,1\,1} & K_{2\,2,2\,2} & K_{2\,2,3\,3} \\
K_{3\,3,1\,1} & K_{3,3,2\,2} & K_{3\,3,3\,3} \\
\end{pmatrix},
\end{equation}
where the indices $1$, $2$, and $3$ refer to the three branches obtained from diagonalization of the 1Q manifold, arranged for increasing value of the energy. Once three independent elements $K_{1\,1,3\,3}$, $K_{1\,1,2\,2}$, and $K_{2\,2,3\,3}$ are fixed, $K$ is fully determined by detailed balance and probability conservation. Specifically, $K$ satisfies
\begin{equation}
\begin{split}
    &K_{e\,e,e'e'} >0 \quad \mathrm{for~} e=e', \\
    &K_{e\,e,e'e'} <0 \quad \mathrm{for~} e\neq e',\\
    &\sum_e K_{e\,e,e'e'} =0,\\
    &\frac{K_{e\,e,e'e'} }{K_{e'e',e\,e} }=e^{\frac{-\hbar\omega_{ee'}}{k_b T'}},
\end{split}
\end{equation}
where $k_b$ is the Boltzmann constant and $T'$ is the temperature. Following the derivation in \cite{abramavicius_coherent_2009}, the solution of Eqs.~(\ref{eq: Pauli}) and (\ref{eq: Liouville}) can be combined in a common Green function
\begin{equation}\label{eq: solution_Pauli}
\begin{split}
    \mathcal{G}_{e_4e_3e_2e_1}(t)&=\delta_{e_4e_3}\delta_{e_2e_1}\theta(t)\sum_p \chi_{e_4p}^{(R)}D^{-1}_{pp}e^{-\lambda_p}t\chi_{pe_2}^{(L)} \\
    &+ (1-\delta_{e_4e_3})\delta_{e_4e_2}\delta_{e_3e_1}\theta(t)\exp{\left[-i\omega_{e'e}t-\frac{1}{2}(K_{e\,e,e\,e}+K_{e'e^{\prime}, e'e'})t-\gamma_{e'e}t\right]}
\end{split}
\end{equation} 
that satisfies $\rho_{e_4 e_3}(t)=\sum_{e_2,e_4}\mathcal{G}_{e_4e_3e_2e_1}(t)\rho_{e_2e_1}(0)$. In Eq.~(\ref{eq: solution_Pauli}), $\lambda_p$ are the eigenvalues of $K$, $\chi^{(R)}$ ($\chi^{(L)}$) is a matrix whose columns (rows) are the right-eigenvectors (left-eigenvectors) of $K$, and $D=\chi^{(L)}\chi^{(R)}$. 
The Green function in Eq.~(\ref{eq: solution_Pauli}) can then be included in the SoS expressions of the MCDS signal by modifying the ESE and ESA contributions. Eqs.~(\ref{eq: SoS-SE}) and (\ref{eq: SoS-ESA}) become
\begin{subequations}\label{eq: SoS-relax}
    \begin{align}\label{eq: SoSrelax-GSB}
    \begin{split}
        &S^{(3)}_{ii}(\Omega_{em},T, \Omega_{abs})\propto \\&+\sum_{e_1,e_2,e_3,e_4}\mathcal{I}m \frac{\mu_{e_1g}(\hat{\epsilon}_1)\mu_{e_2g}(\hat{\epsilon}_2)\mu_{ge_3}(\hat{\epsilon}_3)\mu_{ge_4}(\hat{\epsilon}_4) e^{-i(\omega_{e'e}-i\gamma_{e'e})T}\E_{LO}^*\E_3\E_2\E_1^*}{(\Omega_{em}-\omega_{e_4g}+i\gamma_{e_4g})(\Omega_{abs}+\omega_{e_1g}+i\gamma_{e_1g})}\mathcal{G}_{e_4,e_3,e_2,e_1}(T),
        \end{split} \\ \label{eq: SoSrelax-ESA}
           \begin{split}
        &S^{(3)}_{iii}(\Omega_{em},T, \Omega_{abs})   \propto 
        \\&-\sum_{e_1,e_2,e_3,e_4,f}\mathcal{I}m\frac{\mu_{e_1g}(\hat{\epsilon}_1)\mu_{e_2g}(\hat{\epsilon}_2)\mu_{fe_4}(\hat{\epsilon}_3)\mu_{e_3f}(\hat{\epsilon}_4) e^{-i(\omega_{e'e}-i\gamma_{e'e})T}\E_{LO}^*\E_3\E_2\E_1^*}{(\Omega_{em}-\omega_{fe_3}+i\gamma_{fe_3})(\Omega_{abs}+\omega_{e_1g}+i\gamma_{e_1g})}\mathcal{G}_{e_4,e_3,e_2,e_1}(T).
        \end{split}
    \end{align}
\end{subequations}
Since the GSB contribution in Eq.~(\ref{eq: SoS-GSB}) does not involve excited-state populations during $T$, it is not affected by the relaxation dynamics described in this section.

In this work, population effects are only included in the simulations at fixed detuning $\Delta=+0.62$~meV, reported in Fig. 3 of the main text, to evaluate the overall agreement between the experimental and calculated MDCS spectra. We found the best qualitative agreement with the experimental data for $K_{1\,1,3\,3}^{-1}=0.1$~ps, $K_{1\,1,2\,2}^{-1}=20$~ps and $K_{1\,1,3\,3}^{-1}=0.8$~ps. While we do not expect relaxation to impact on the spectral peak positions, considering additional couplings to the phonon bath, excitonic transport, cavity losses, and other coherent and incoherent mechanisms might be necessary to fully describe the relative intensities of the polariton peaks \cite{Mondal2023}.

\section{Analysis of the B feature}
The linear prediction by singular value decomposition (LPSVD) algorithm as implemented for data acquired by NC-MDCS \cite{swagel_analysis_2021} considers slices of the complex 2D spectrum $S_{1Q}(-\omega_\tau,T,\omega_t)$ along the absorption energy axis. For each column, the algorithm analyzes $\tau$ domain data, finding the largest 12 ``poles,'' each of which has an amplitude, phase, absorption frequency $\omega_\tau$, and damping rate. These poles are then filtered to exclude spurious poles that are small in amplitude or that have $\omega_\tau$ outside the bandwidth of the laser spectral bandwidth. From the lists of poles found at each $\omega_t$, we can construct a 2D spectrum. One benefit of this approach is that the poles can then be separated into groups by their absorption and emission frequency coordinates. Most poles correspond to the LP or UP diagonal or off-diagonal peaks. Note that because of the non-Lorentzian lineshape of the data, multiple poles generally contribute to each peak. For the XXXX and -- -- + + polarizations, some poles correspond to the B feature. The constructed spectrum using only the B poles is used to identify the absorption and emission energy coordinates and the magnitude of the B feature shown in Fig.~2 of the main text.

The signal-to-noise ratio of the B peak in the C-MDCS dataset at negative $\Delta$ was sufficiently high to allow analysis of the MDCS spectra before averaging multiple acquisitions. Also, for $\Delta < 0$ the B peak is well separated from the LP feature. Together these factors enabled us to directly fit the line slices in the frequency domain and determine the peak positions. For each detuning, the maximum value of the 2D signal in the region of the diagonally elongated B peak was determined and its position was used to extract the anti-diagonal line cut of the 2D spectrum passing through the peak. The line cuts were then fitted using Voigt profiles. The coordinates of the center of the fitted lineshapes were projected back onto the emission and absorption axes to obtain the peak energies displayed in Fig.~2 of the main text.

\begin{figure}[h!]
\includegraphics[width=0.8\textwidth]{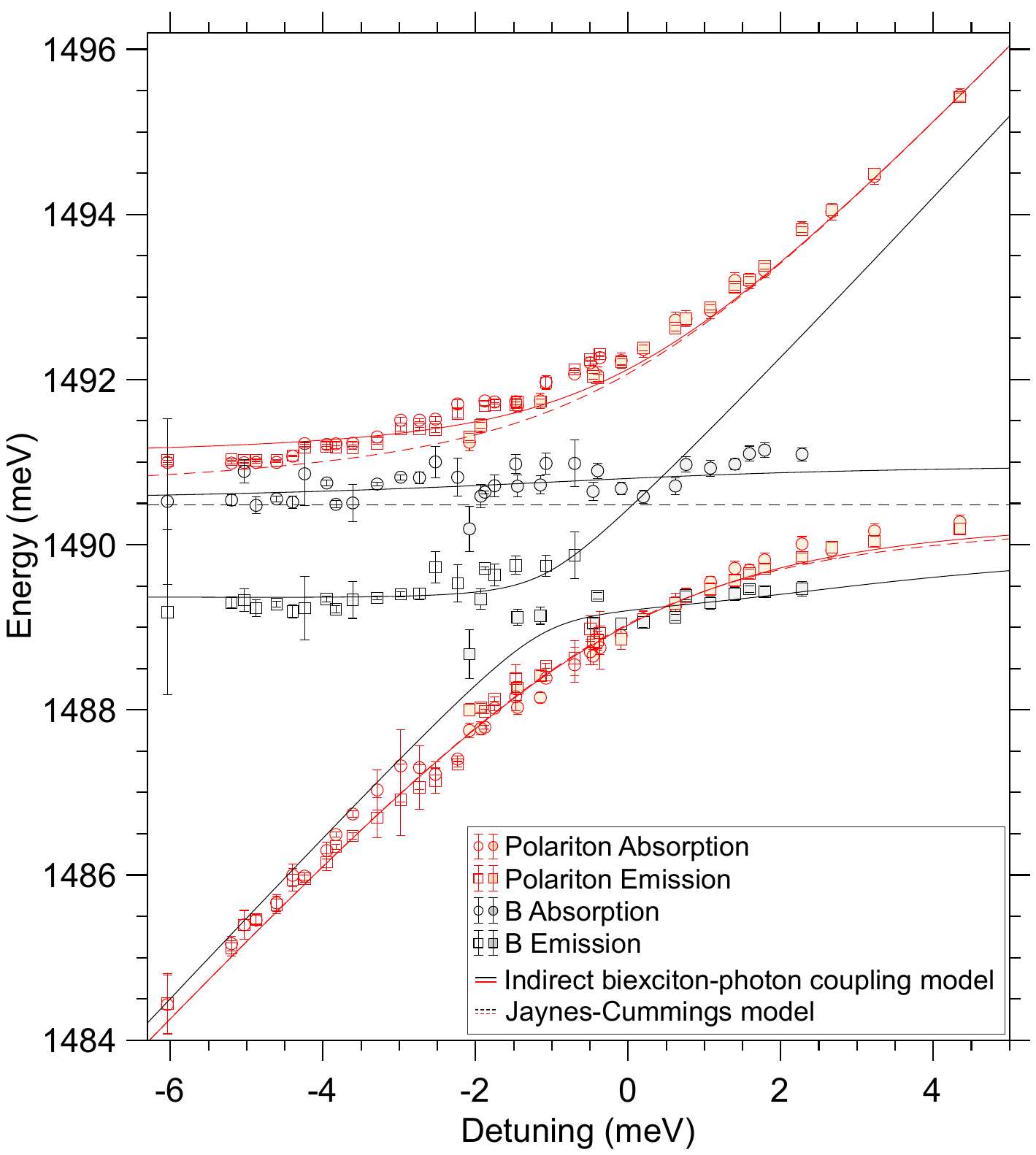}
\caption{Extracted absorption (circles) and emission (squares) energies from both NC-MDCS (filled) and C-MDCS (open) for the UP, LP, and B features as a function of $\Delta$. Dashed lines show fits to Eq.~(1) and a fixed (dark) exciton energy, and solid lines show fits to the indirect biexciton-photon coupling model. Colors highlight branches related to the UP or LP (red) and to B (black). Error bars indicate 95 \% confidence intervals.}
\label{Fig: uncertainty}
\end{figure}

The uncertainty of the spectral coordinates obtained from the individual fits is generally much smaller than the scatter of the points in Fig.~2. Consequently, the uncertainty is not related to the noise within individual spectra. We distinguish between two different noise sources that impact the extraction of the B peak: (i) variation of spectra acquired under otherwise identical conditions, (ii) lineshape distortions due to the interference with the spectrally overlapping tails of the stronger LP peak. For the C-MDCS data, the latter is not relevant since the B and LP peaks are spectrally separated at the considered detunings. To estimate the uncertainty related to (i), we acquired multiple spectra at the same value of $\Delta$, fitted them separately and analyzed the statistical distribution of the results. This allowed us to calculate an uncertainty estimate (95 \% confidence interval) for those data points. In the NC-MDCS data, we could not analyzed spectra before averaging multiple acquisition, due to the limited signal-to-noise ratio and long acquisition time. We then focused on noise (ii), by adopting a non-parametric bootstrapping method to estimate this source of uncertainty. At each detuning, we considered a sample of $n_p=$30 anti-diagonal linecuts passing through a point $p_i$ with $i=1,\dots n_p$ in the proximity of the peak position retrieved by the LPSVD. The distance from the retrieved peak position was randomly extracted within the linewidth of the B peak. We fitted each linecut separately with Voigt profiles and calculate the mean value of the peak positions obtained from the fits. We repeated the procedure, re-sampling from the initial collection of points $p_1\dots p_{n_p}$, for 350 iterations. Finally, we analyzed the distribution of the results across the iterations and calculated an uncertainty estimate (95 \% confidence interval). This method allowed us to estimate the error in the retrieved peak positions, in particular along the spectral dimension not directly considered in the LPSVD. A version of Fig. 2 that includes error bars generated as detailed above is shown in Fig.~\ref{Fig: uncertainty}.

\section{Model fit}
We used the eigenvalues obtained by diagonalizing Eq.~(2) in the main text, corrected by the background interaction as determined by Eq.~(\ref{eq: backgroundInt}), to globally fit the energies their polaritons and B peaks from the NC- and C-MDCS datasets, extracted by using the procedures described above. The best fit of the parameters are reported in Table \ref{tb: globalfit} and used to compute the solid lines in Fig.~2 of the main text. The energy of the dark exciton is assumed to be equal to that of the bright one: $\Edark=\EX$. 
\begin{table}[h]
\caption{\label{tb: globalfit}
Parameters from the fit of the NC- and C-MCDS absorption and emission energies.}    
\begin{ruledtabular}
\begin{tabular}{ cccccccc }
Parameter & 
$\EX$ & 
$\Delta E_{b}^{B}$ & 
$\ORabi$ &  
$\Obiexc$ & 
$g_{\mathrm{spin}}$ & 
$g_{B}$ &  
$\gamma_{B}$ 
\\
{Value (meV)}& {1490.48}&  {0.86}& {3.03}& {0.75}& {0.75}&  {-1.01}& {1.16}\\
\end{tabular}
\end{ruledtabular}

\end{table}

We estimated the homogeneous dephasing and and the inhomogeneous broadening to the values reported in Table \ref{tb: gammafit}. To extract $\gamma_X$, we fitted the antidiagonal slice crossing the B peak at $\Delta\approx -2$~meV, where the peak is mostly excitonic, using a Voigt profile. The Lorentzian width obtained from the fit was then used as $\gamma_X$ in the simulation. The other broadening parameters and the EID and EIS factors have been selected by comparison between the peak widths of the experimental spectra and those simulated by using Eqs.~(\ref{eq: SoS}), seeking the best qualitative match of the experimental linewidths and amplitudes.

\begin{table}[h]
\caption{\label{tb: gammafit}
Values of the broadening parameters used in the simulations.}
\begin{ruledtabular}
\begin{tabular}{ cccccccc}
Parameter & 
$\gamma_C$ & 
$\gamma_{X}$ &
$\gamma_{D}$ & 
$b_{X}$ & 
$b_{D}$ &
$\Delta{\gamma}^{\mathrm{(EID)}}$ & 
$\Delta E^{\mathrm{(EIS)}}$ 
\\
{Value (meV)}&  {0.6}& {0.1}& {0.1}& {0.5}& {0.1} &{0.1} & {-0.3}\\
\end{tabular}
\end{ruledtabular}
\end{table}
Finally, we fitted the amplitude of the B peak in the NC-MDCS dataset using Eqs.~(\ref{eq: SoS}), to obtain the values of the transition dipole strengths, keeping all the other parameters fixed. Separate parameters for the $g\to 1Q$ and $1Q \to 2Q$ transitions have been used to account for saturation of the 2Q manifold. 
The results are reported in Table \ref{tb: mufit}. The parameters reported in Tables \ref{tb: globalfit}, \ref{tb: gammafit} and \ref{tb: mufit} were used for the simulations reported in Fig.~3 of the main text. 
\begin{table}[h]
\caption{\label{tb: mufit}
Dipole strengths obtained from fitting the normalized amplitude of B.}
\begin{ruledtabular}
\begin{tabular}{ ccccc}
Parameter & 
$\mu_{qm}^{1Q}$ &
$\mu_{qm}^{2Q}$ & 
$\mu_{D}^{1Q}$ & 
$\mu_{D}^{2Q}$ 
\\
{Value} & {0.3}& {0.29}& {0.04}& {0.25}\\
\end{tabular}
\end{ruledtabular}
\end{table}

\newpage

\section{Rephasing 1Q spectra for different detunings}

\begin{figure}[h!]
\includegraphics[width=0.9\textwidth]{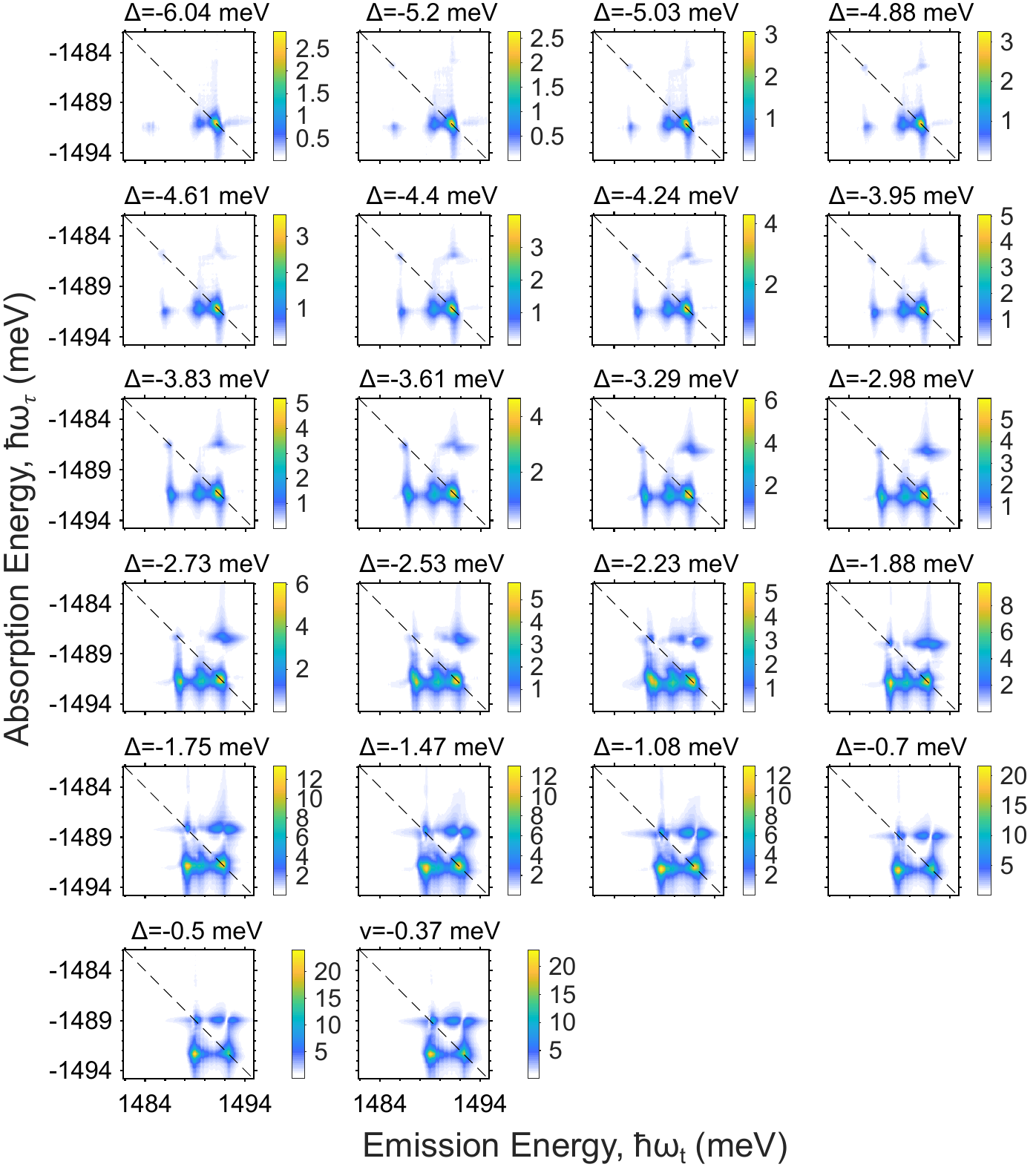}
\caption{C-MCDS 1Q rephasing spectra obtained with -- -- + + polarization for the detuning values $\Delta$ analyzed in the main text. }
\label{Fig: BF dataset}
\end{figure}

\begin{figure}[h!]
\includegraphics[width=0.9\textwidth]{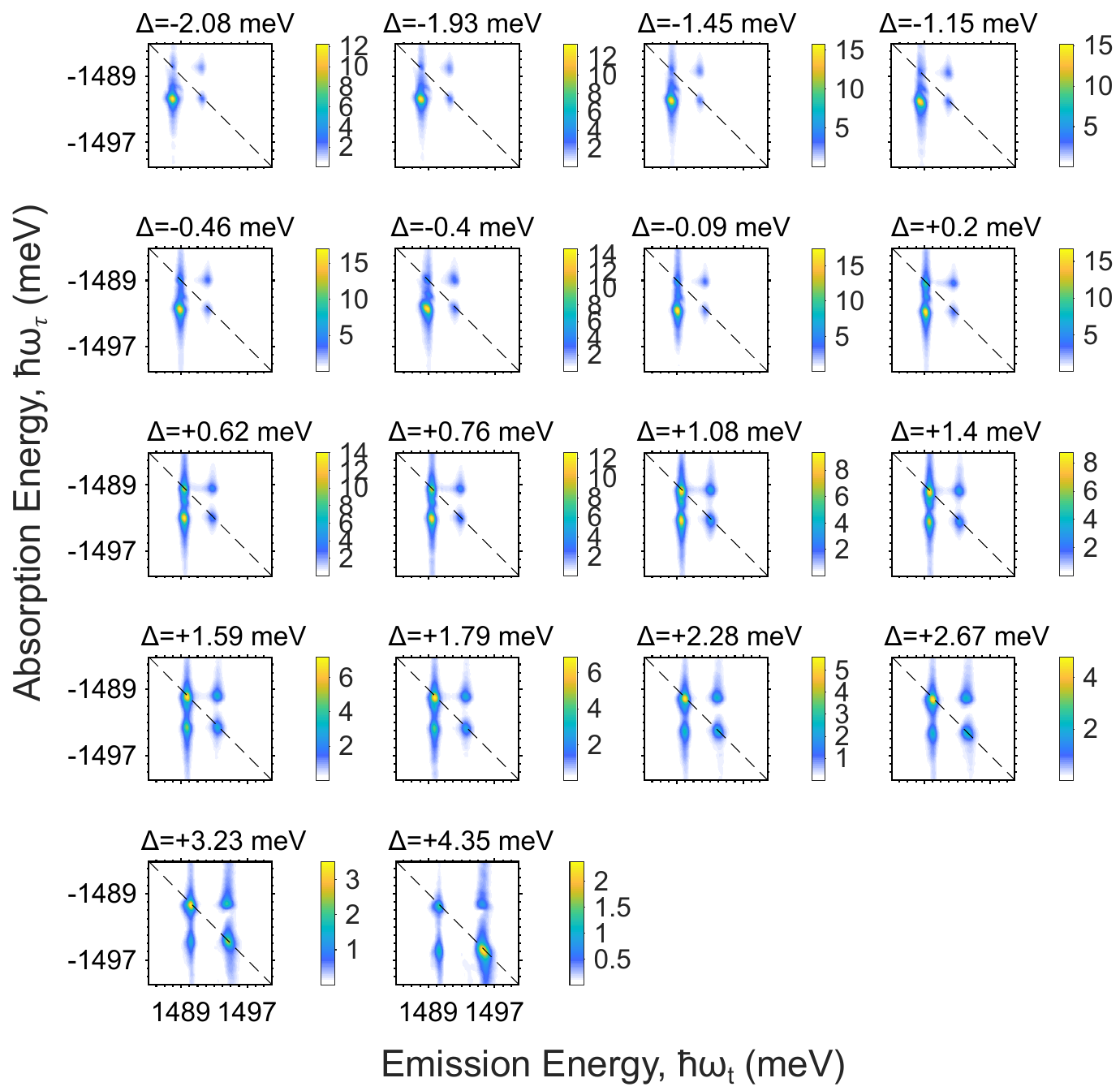}
\caption{NC-MCDS 1Q rephasing spectra obtained with + + -- -- polarization for the detuning values $\Delta$ analyzed in the main text.}
\label{Fig: MONSTRspectra}
\end{figure}

Figure \ref{Fig: BF dataset} and \ref{Fig: MONSTRspectra}, acquired using the C-MDCS  and the NC-MDCS setups respectively, show magnitude of the 1Q spectra for + + -- -- polarization as a function of detuning.

\newpage
\section{Dependence of 1Q C-MDCS spectra on the incident angle}
\begin{figure}[h!]
\includegraphics[width=\textwidth]{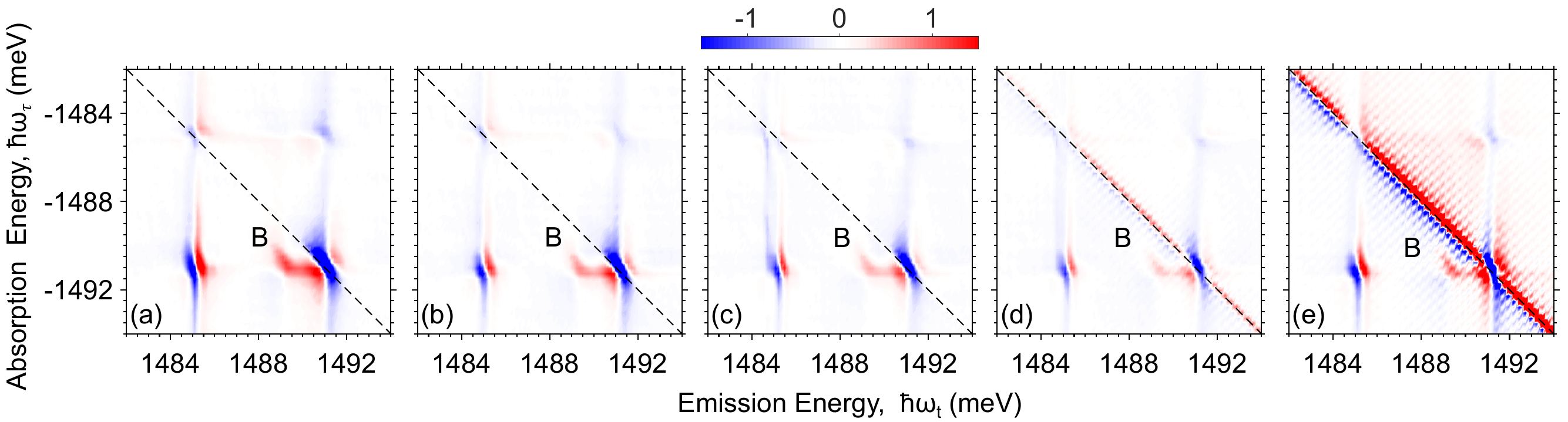}
\caption{Real part of the 1Q rephasing spectra at $\Delta \approx -$5~meV obtained in the C-MDCS geometry with XXXX polarization as a function of the incident angle $\theta_i$. The B peak is highlighted by the label B. From left to right: (a) $\theta_i=$ 5.14$^\circ$, (b) 2.63$^\circ$, (c) 1.07$^\circ$, (d) 0.58$^\circ$ and (e) 0$^\circ$. }
\label{Fig: angle}
\end{figure}

Fig.~\ref{Fig: angle} shows the real part of 1Q rephasing spectra obtained with C-MDCS at fixed detuning and varying the common angle of incidence of the excitation pulses on the sample ($\theta_i$). The B peak is observed at all angles within the measured range. Because the B feature is shifted off of the diagonal, its value can still be extracted near $\theta_i=0^\circ$ where a large linear artifact is present.

\section{2Q NC- and C-MDCS spectra}
\begin{figure}[h!]
\includegraphics[width=.5\textwidth]{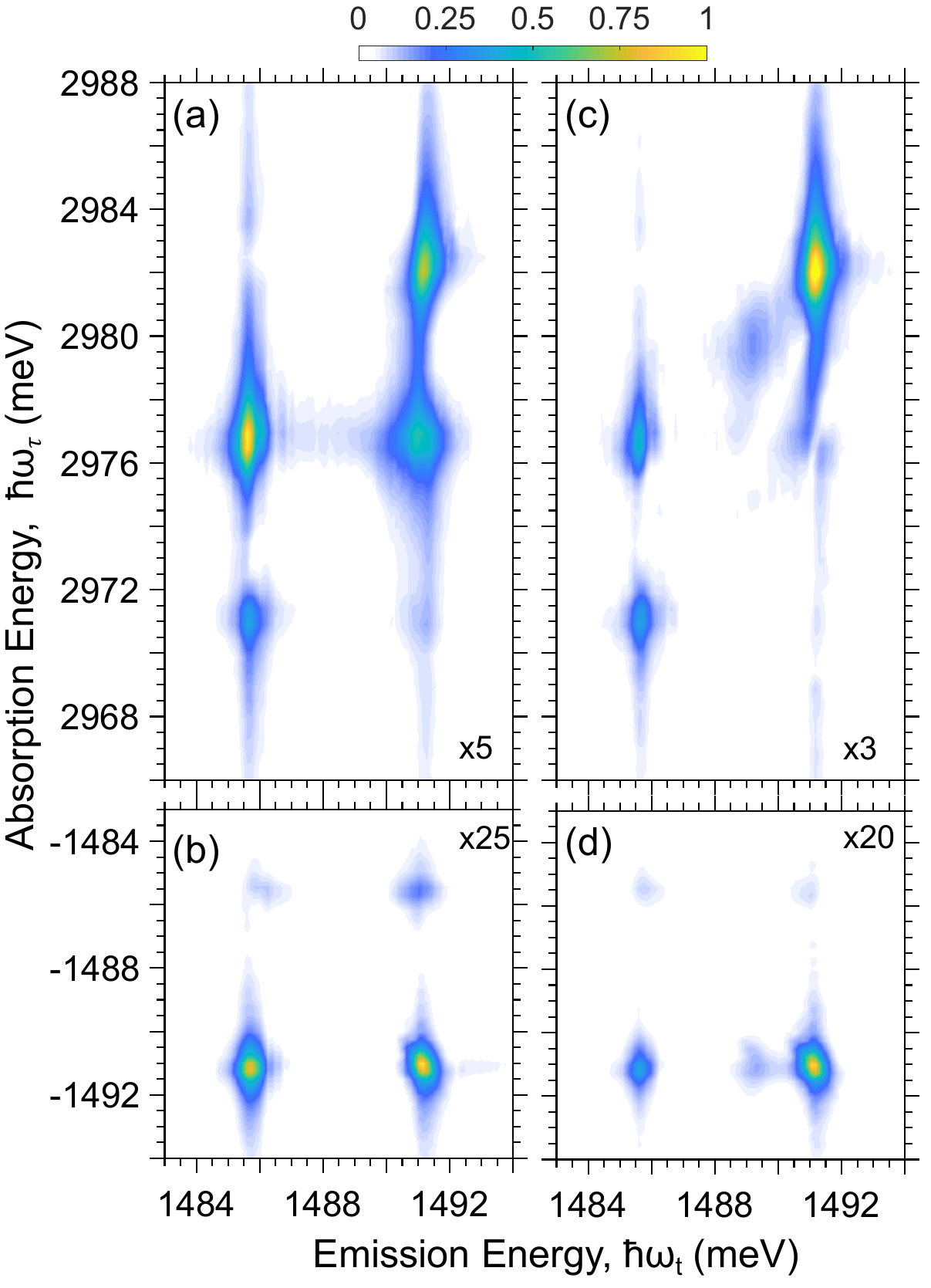}
\caption{C-MDCS 2Q spectra at $\Delta \approx -$4~meV obtained in the C-MDCS geometry with + + + + (a) and XXXX (c) polarizations. The corresponding spectra for the 1Q rephasing data are reported in the panels below: (b) is + + + + polarization and (d) is XXXX polarization. Each spectrum was normalized to its highest value and the scale factor is reported.}
\label{Fig: 2Qspectra}
\end{figure}

To confirm the biexcitonic origin of the B feature, we measured 2Q spectra, $S_{2Q} (\tau, \omega_T,\omega_t)$, which are directly sensitive to excitations of two-particle resonances \cite{karaiskaj_two-quantum_2010}. Results are shown in Fig.~\ref{Fig: 2Qspectra} for NC- and C-MDCS. Similarly to 1Q spectra, excitation under co-linear polarization produces an additional feature besides the ones observed also in co-circular polarized conditions and corresponding to the two-upper (2UP), two-lower (2LP) and mixed polariton (2MP) peaks of the doubly excited polariton manifold \cite{Autry2020}. The emission energy of this feature is the same as the one of the B peak in the 1Q spectrum, while its absorption energy  $\hbar \omega_t=2980 \mathrm{\,meV}\approx2\EX -\Delta E_{b}^{B}$ agrees with the 2Q energy of B upper branch.

\input{si.bbl}